\documentclass{article}
\usepackage{waspaa19,amsmath,amsfonts,graphicx,url,times, float, subfig}
\usepackage{graphicx}
\usepackage{mathtools, multirow, arydshln}
\usepackage{booktabs}
\usepackage{pst-node}

\usepackage{color}

\title{Deep Tensor Factorization for Spatially-Aware Scene Decomposition}

\twoauthors
 {Jonah Casebeer*, Michael Colomb*\thanks{* These two authors contributed equally}}
    {University of Illinois at Urbana-Champaign\\
Computer Science Dept.\\
    jonahmc2, colomb2@illinois.edu}
 {Paris Smaragdis$^\dagger$\thanks{$\dagger$ This work is supported by NSF grant \#1453104}}
    {University of Illinois at Urbana-Champaign\\
Computer Science Dept.\\
    Adobe Research}

\begin{document}
\ninept
\maketitle
\begin{sloppy}
\begin{abstract}
We propose a completely unsupervised method to understand audio scenes observed with random microphone arrangements by decomposing the scene into its constituent sources and their relative presence in each microphone. To this end, we formulate a neural network architecture that can be interpreted as a nonnegative tensor factorization of a multi-channel audio recording. By clustering on the learned network parameters corresponding to channel content, we can learn sources' individual spectral dictionaries and their activation patterns over time. Our method allows us to leverage deep learning advances like end-to-end training, while also allowing stochastic minibatch training so that we can feasibly decompose realistic audio scenes that are intractable to decompose using standard methods. This neural network architecture is easily extensible to other kinds of tensor factorizations.
  
\end{abstract}


\section{Introduction}
\label{sec:intro}
A primary goal of audio scene decomposition is to identify the individual audio sources that constitute a scene, as this is a precursor step for many tasks in audio processing like denoising, source separation, audio scene remixing, and source localization. In this paper we develop a neural network architecture that can be interpreted as a non-negative tensor factorization and then later use it to identify the unique sources of a scene in an unsupervised manner. In particular we're interested in realistic audio scenes observed with multiple microphones that record various perspectives of the scene. For example, this could be the modern living room where multiple cellphones, a laptop, smart speakers and other smart devices are present. We aim to deploy a model that can leverage such an acoustically diverse set of microphones even without labels or prior knowledge to construct a set of frequency and channel dictionaries, whose components inherently contain useful information about the scene, enabling further processing. We call our model a Deep Non-Negative Tensor Factorization (DNTF). This model is highly flexible in that it can easily be adapted to leverage modern deep learning advances, like complex nonlinear transforms, stochastic training, adaptive front ends, arbitrary regularizers, and specialized loss functions. It can also be adapted to represent other kinds of tensor factorizations that incorporate varying types of tensor products (e.g. Kronecker, convolutional, or recurrent).

Models for decomposing single channel scenes usually rely on a supervised training phase, during which they are explicitly pointed towards sources in order to build models for them. Efforts to decompose a scene without this prior learning step have not been as successful, as pairing extracted components with specific sources without additional information is difficult. Multi-channel settings address this problem by leveraging inter-channel information that can facilitate the pairing process. These methods have been used for separation and remixing based on user input and without labels \cite{ozerov2011multichannel, ozerov2010multichannel}. Similar work like \cite{sawada2012efficient} clusters components based on other statistical information. Non-tensor approaches like \cite{seichepine2014soft} run separate decompositions on each channel with soft parameter sharing constraints. Recently, this class of factorization techniques has started to fall under the deep learning umbrella. These methods interpret non-negative autoencoders as an extension of non-negative factorization methods \cite{smaragdis2017neural, venkataramani2017neural}. These hybrid models benefit from the advances in deep learning while maintaining the interpretability that many deep models forgo. In particular, the parameters of a trained model can convey information about the decomposed scene, the nature of the constituent sources, and their activation patterns.

Deep models that are not based on factorizations have been explored for multi-channel audio tasks like denoising and source separation. Some models assume a specified microphone geometry and attempt to predict the weights of a beamformer in either time or frequency \cite{xiao2016deep, li2016neural}. We focus on scenes similar to those studied by \cite{casebeer2018multi}, where very few assumptions are made about the microphone locations, arrangement, and synchronization. However, these techniques are supervised, and thus require labeled training datasets. More recently, \cite{tzinis2019unsupervised} explored methods for training a single channel source separation system on stereo audio recordings. We work towards a model that can leverage these architectural advances without being as opaque as generic deep approaches such that its parameters contain readily interpretable information about the scene.

In our work we aim to extend \cite{smaragdis2017neural} to tensors of arbitrary orders. Using such tensors allows us to decompose scenes made up of many microphones with a single factorization. Then, we use the parameters of our model to reconstruct the audio sources while never relying on labeled data. Our model preserves the interpretability of previous tensor methods while leveraging modern advances in deep neural network training. The parameters learned by the model could be used for downstream tasks like source identification, spatial scene remixing, and single and multi-channel source separation. Decomposing a scene is independent of the downstream task. This means that rerunning and optimizing those tasks, which are inherently simpler now, can be performed much easier as compared to directly completing the task from scratch.

Demo available at: \textit{https://jmcasebeer.github.io/projects/dntf/}

\section{Methods}
We now describe the model that we employ in this paper. This is a reformulation of the PARAFAC tensor factorization as a non-negative tensor factorization performed by a neural network. We also develop an application of the model to blind source separation.
\label{sec:model}


\subsection{The DNTF Model}
DNTF is built on the PARAFAC \cite{harshman1970foundations} formulation of 3-order tensor factorization, in which a tensor $\mathbb{X}^{C \times F \times T}$ is factored into the matrices $\mathbf{D}^{C \times K}$, $\mathbf{W}^{F \times K}$, $\mathbf{H}^{T \times K}$ such that:
\begin{align}
\label{eq:parafac}
    x_{c, f, t} = \sum \limits_{k=1}^{K}d_{c,k}w_{f,k}h_{t,k}
\end{align}
where $K$ typically $\ll C, F, T$ is a hyperparameter that defines the shared dimension of the matrix factors.

Note that in this form, $\mathbb{X}$ cannot be represented with a matrix product over its factors, so we define $\mathbf{X}^{T \times CF }$ as
a flattened $\mathbb{X}$. In \cite{bro1998multi} it was shown that:
\begin{align}
\label{eq:matrix_parafac}
    \mathbf{X} = \mathbf{H} \cdot (\mathbf{D} \diamond \mathbf{W})^\top
\end{align}
where $\diamond$ refers to the Khatri-Rao product. Building on the work from \cite{smaragdis2017neural}, we interpret Eq. \ref{eq:matrix_parafac} as the decoding step in a deep autoencoder, where $\mathbf{H}$ is the embedding. To ensure a non-negative factorization of $\mathbf{X}$, we define $\mathbf{D} = \sigma_D(\mathbf{\widetilde{D}})$ and $\mathbf{W} = \sigma_W(\mathbf{\widetilde{W}})$, where $\mathbf{\widetilde{D}}^{C \times K}$ and $\mathbf{\widetilde{W}}^{F \times K}$ are the weight matrices of the decoder and $\sigma$ are non-negative functions mapping $\mathbb{R}^{\{C,F\} \times K} \to \mathbb{R}_{\geq 0}^{\{C,F\} \times K}$. 

Then to define the encoding step, we need to be able to invert $\mathbf{D} \diamond \mathbf{W}$. Since directly computing the pseudo-inverse $\mathbf{D}^{+} \diamond \mathbf{W}^{+}$ is intractable during training, but we know the inverse operation is a Khatri-Rao product of matrices with known shape, it can be approximated as:
\begin{align}
    (\mathbf{D} \diamond \mathbf{W})^{+} \ = \ \mathbf{D}^+ \diamond \mathbf{W}^+ \ \approx \ \mathbf{D}^\ddagger \diamond \mathbf{W}^\ddagger
\end{align}
where ${\mathbf{D}^\ddagger}^{K \times C}$ and ${\mathbf{W}^\ddagger}^{K \times F}$ are the weight matrices of the encoder. To guarantee the non-negativity of $\mathbf{H}$, the output of the encoder must pass through a non-negative activation function $\sigma_H$. While this nonlinearity obfuscates the relationship between the encoder and decoder, as long as an appropriate activation function is chosen, such as ReLU or softplus, its effect becomes insignificant as the model performance improves. And thus, the DNTF model is defined as:
\begin{align}
    \textbf{Encoder:} && \mathbf{H} &=  \sigma_H (\mathbf{X} \cdot ( {\mathbf{D}^\ddagger}^\top \diamond {\mathbf{W}^\ddagger}^\top)) \\
    \textbf{Decoder:} && \hat{\mathbf{X}} &= \mathbf{H} \cdot ( \sigma_D (\mathbf{\widetilde{D}}) \diamond \sigma_W(\mathbf{\widetilde{W}} ))^\top
    \label{eq:autoencoder}
\end{align}
In the case of multi-channel audio scene decomposition, $\mathbb{X}^{C \times F \times T}$ is the tensor containing the short-time Fourier transform (STFT) over the observed multi-channel audio, where $C$ is the number of channels or microphones, $F$ is the number of frequency bins, and $T$ is the number of short-time analysis windows. Thus, $\mathbf{H}$ contains the component activations over time, $\mathbf{D}$ contains the distribution of each component over the channels, and $\mathbf{W}$ is the spectral dictionary over the components. Fig. \ref{fig:model} shows the results of training DNTF over a simple, artificial audio scene in order to build some intuition. To verify the correctness of the decomposition at a particular channel, time, and frequency, sum over the products of the corresponding channel, frequency and activation values for each component as defined in Eq.\ref{eq:parafac}.
\begin{figure}[ht]
    \centering
    \hspace*{-4mm}\includegraphics[scale=0.45]{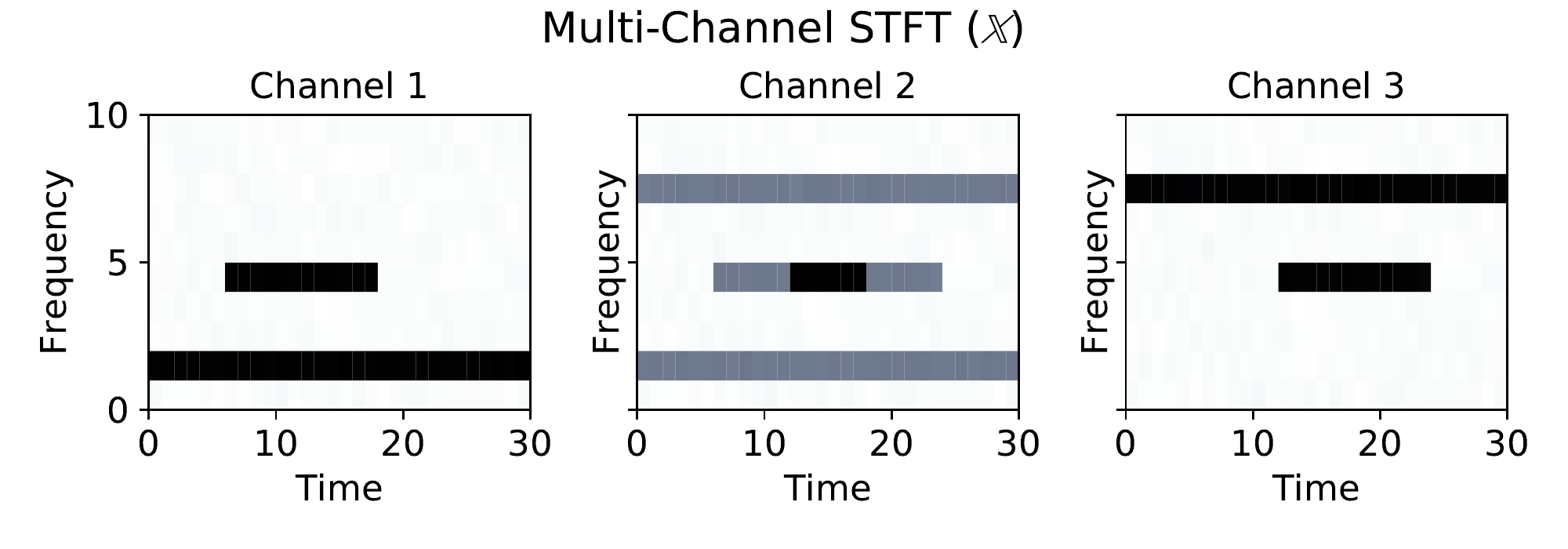}
    \hspace*{-4mm}\includegraphics[scale=0.45]{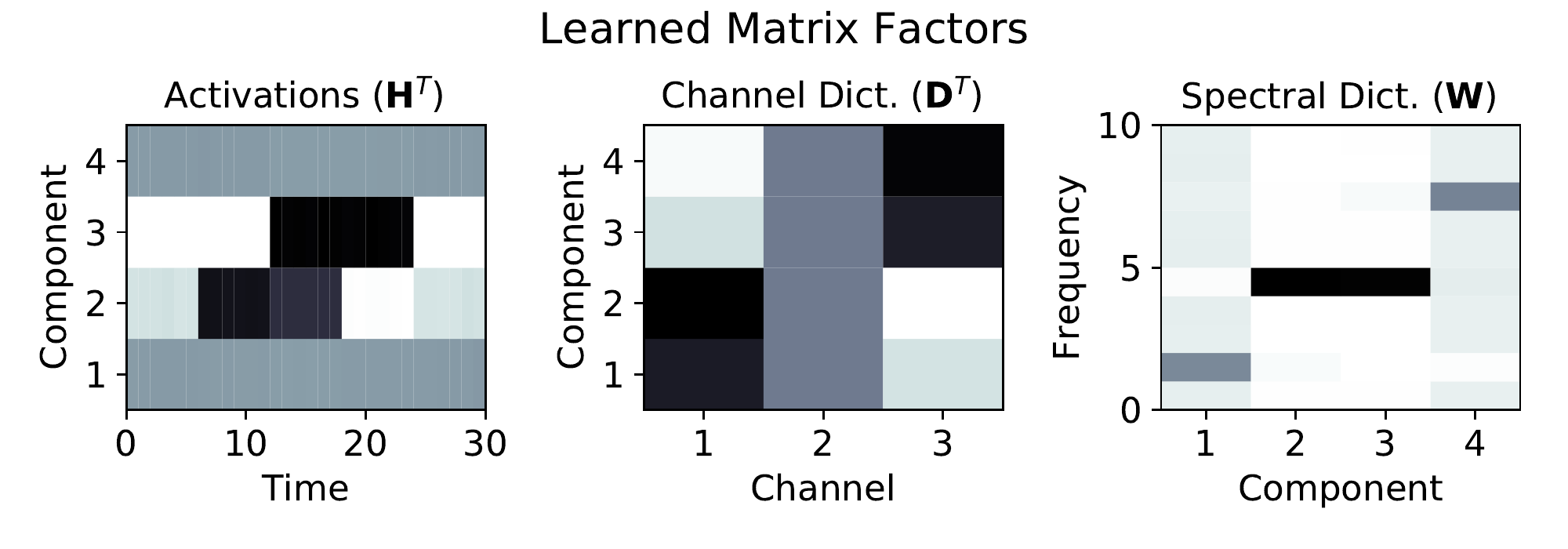}
    \caption{Example scene decomposition. \textbf{Top row:} the simulated STFT over each channel ($\mathbb{X}$). The tensor, shown with a separate plot for each channel, contains two true sources. The first source is strongest in channel 1, while the second is strongest in channel 3. Channel 2 exhibits a mixture containing both. \textbf{Bottom row:}
    the learned matrix factors. By inspection, the channel dictionary's components form two distinct groups. Examining the first group's components in the spectral dictionary shows they match the source in channel 1, while the second group's components match the source in channel 3. The channel dictionary uses all components equally in channel 2 since both sources are active in it. This shows that the model can decompose a scene into its constituent sources in an interpretable fashion.}
    \label{fig:model}
    \vspace{-5mm}
\end{figure}
\subsection{Loss and Training}
We train DNTF as an autoencoder, where we define the element-wise reconstruction loss as the Itakura-Saito divergence \cite{fevotte2009nonnegative}:
\begin{align}
    d(x_{c,f,t} \mid \hat{x}_{c,f,t}) = \frac{x_{c,f,t}}{\hat{x}_{c,f,t}} - \log{\frac{x_{c,f,t}}{\hat{x}_{c,f,t}}} - 1
    \label{eq.isdiv}
\end{align}

We train our model stochastically by randomly selecting minibatches from $\mathbb{X}$ of length $M$ over the time dimension, such that at each iteration during training we are only encoding and decoding a tensor of size $(C \times F \times M)$. By never loading the entire scene into memory or performing computation over the entire scene, stochastic training enables the model to decompose audio scenes that are far too large for standard factorization algorithms, and also allows us to learn a factorization in an online manner, observing only one time frame at a time.

Also, because DNTF learns an encoder as opposed to explicit activations over the training data, it can directly encode new observations from the scene. Traditional factorization algorithms, including those with stochastic variants, generally rely on further iterative training once a new input is provided.





\subsection{Application to Blind Source Separation}
Given a single channel NMF decomposition, it is difficult to identify which components belong to which source, especially if the sources have similar spectra. We leverage the channel dictionary $\mathbf{D}$ learned by the DNTF to group the frequency components in $\mathbf{W}$ by assuming that different sources are spatially separated. With this restriction, if two frequency components $\mathbf{w}_i$ and $\mathbf{w}_j$ have similar channel components $\mathbf{d}_i \approx \mathbf{d}_j$, then they likely come from the same source. Thus, we can pose this problem of identifying $N_S$ sources as an unsupervised clustering problem, where each cluster represents a source. We run k-means clustering with $N_S$ centers on $\mathbf{D}$ where each of the $K$ components is a data-point of dimension $C$. To reconstruct the sources, we propose two methods. The first leverages k-means cluster assignments, and the other k-means cluster centers.

\subsubsection{Cluster Assignment Based Separation}
 By interpreting the k-means cluster assignments as source assignments we can reconstruct the source corresponding to a particular k-means cluster. First, select a cluster. Then, run Eq. \ref{eq:autoencoder} where any component not in the cluster is set to zero. Stack the multi-channel reconstructions for each source into a multi-source multi-channel tensor $\mathbb{S}^{N_S \times C \times F \times T}$. We recover a single-source single-channel recording by using $\mathbb{S}$ as a Wiener filter. First, recover source $s$ in each channel by using the channel's source estimate as a ratio mask. Then, sum the masked results across channels, i.e. $\sum \limits_{c=0}^{C} \mathbb{X}_{c,,} \frac{\mathbb{S}_{s,c,,}}{\sum_{n=0}^{N_S} \mathbb{S}_{n,c,,}}$ where the division and multiplication operations are element-wise. 
 



\subsubsection{Cluster Center Based Separation}
By interpreting the k-means cluster centers as the distribution of each source over all channels, we can reconstruct each source by constructing a linear system for each STFT short-time frame index $t \in \{1..T\}$ and solving. The system for each short-time frame index $t$ is:
\begin{align}
    \text{Suppose}\quad& \mathbb{X}_{(,,t)} \in \mathbb{R}_{\geq 0}^{C \times F}, \mathbf{C} \in \mathbb{R}_{\geq 0}^{C \times N_S},  \mathbf{S_t} \in \mathbb{R}_{\geq 0}^{N_S \times F} \nonumber \\[.5em]
    &\mathbb{X}_{(,,t)} = \mathbf{C} \cdot \mathbf{S_t} 
\end{align}
where $\mathbf{C}$ is the cluster center matrix whose columns hold the predicted k-means cluster centers, which does not change across $t$. $\mathbf{S_t}$ holds the frequencies emitted by each source at frame index $t$. We solve the above system for $\mathbf{S_t}$ at every short-time frame index $t$ with euclidean NMF where $\mathbb{X}_{(,,t)}$ and $\mathbf{C}$ are fixed. Then, given an $\mathbf{S_t}$ for every $t$, we reconstruct the $s^{\text{th}}$ source by taking the $s^{\text{th}}$ row from every $\mathbf{S_t}$ and stacking them to get a source STFT matrix of size $(F \times T)$. We use Euclidean NMF since the matrix $\mathbf{C}$ is computed with k-means which uses Euclidean distance.







\section{Experiments}
\label{sec:experiments}
We evaluate the model by decomposing a variety of reverberant audio scenes and by attempting to recover the constituent sources by clustering the learned channel dictionary in an unsupervised manner. The same model and hyperparameters are used in every scene to demonstrate that DNTF learns a general decomposition not specific to a particular setup. We measure the performance of the model with the BSS eval v3 metrics Source to Distortions Ratio (SDR), Source to Interferences Ratio (SIR) and Source to Artifacts Ratio (SAR) \cite{vincent2006performance}. We compute these metrics with respect to each scene's single source reverberant simulation. Each setup is simulated with 50 randomly generated source-mic configurations in a 10 by 10 meter room using the \textit{pyroomacoustics} image-source module with second order echoes and an absorption coefficient of 0.85 \cite{scheibler2018pyroomacoustics}. Figure \ref{fig:room} shows an example room.
\begin{figure}
    \centering
    \includegraphics[scale=0.45]{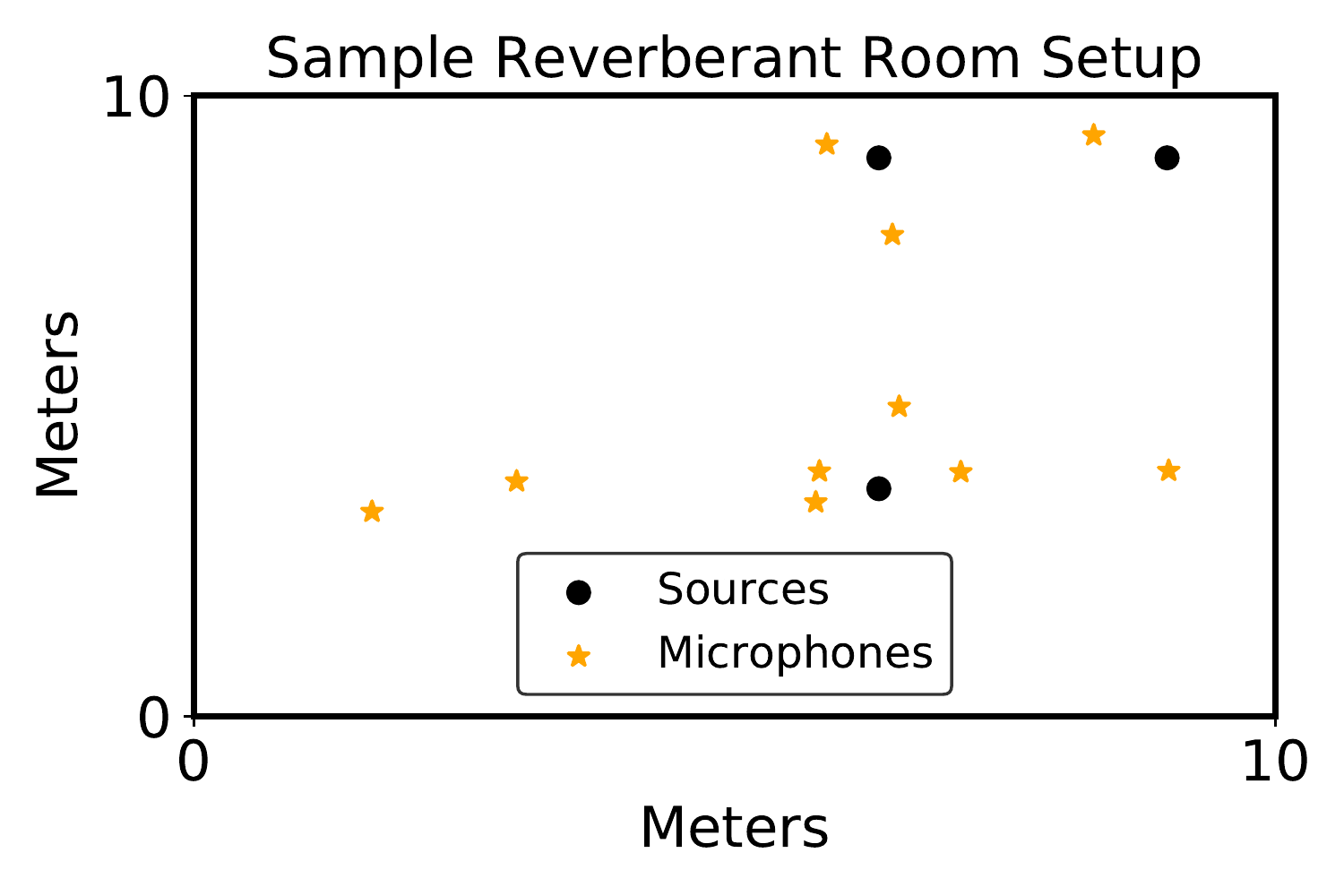}
    \caption{A reverberant 10 by 10 meter room with a random arrangement of three point sources and 10 microphones. Recordings made in this room include up to second order wall reflections.}
    \label{fig:room}
    \vspace{-5mm}
\end{figure}

For each experiment, we randomly select unique speakers from the train fold of the TIMIT dataset \cite{garofolo1993darpa}, placing no restrictions on gender. We then concatenate random recordings from the same speaker to generate audio samples of sufficient length. We require that the closest object to each source is a mic, so to construct a viable configuration, we place each point source randomly on a grid and then place a mic near each. Each source has equal power. More mics are then randomly placed anywhere in the room to total 10 mics. This process guarantees that the sources are spatially distinct and provides a diverse sampling of the room. As for the ambient sources, we use one of "airport", "car", "bus", "subway", "exhibition", "babble", "street", "restaurant", "can-kick", or "drill" from \cite{liu2014experiments}. We simulate all scenes for 10 seconds.

In order to decompose the input scenes in the experiments below, we trained the proposed model using the Itakura-Saito divergence as defined in Eq \ref{eq.isdiv} for the loss function. We trained using a batch size of 15 and trained on 3000 batches. We used the Adam algorithm to update the model parameters, with a learning rate of $10^{-2}$, and decomposed the input to 100 components. The audio inputs were presented as a magnitude STFT with 1024 point DFT frames and a 256 point hop size. When reconstructing the audio we use the phase from that source's loudest channel. The components were clustered using k-means. On an Nvidia GTX 1080, it took ~30 seconds to factor and separate a 10 second scene with 10 channels. We note that we found the models rather robust to different hyperparameters. 




We use three different experiments to demonstrate that DNTF can decompose very different scenes with the same unsupervised model and still recover information about the constituent sources by leveraging the channel dictionary. 
To measure how well the scene was decomposed we measure the SDR, SIR and SAR values of the reconstructed isolated sources.
For each experimental setup we compare cluster assignment based separation and cluster center based separation as defined in the model section. Again, for cluster assignment based separation, we use the estimated magnitude STFTs as Wiener filters.

In the first experiment, we simulate three point sources in a room. This setup tests the model's ability to correctly identify point sources and distinguish between several spatially distinct objects. The results of this experiment are shown in figure \ref{fig:3s_nonoise}. We can see that the resulting separation metrics are quite high on average, on par with supervised methods. Here we see the center based method outperforming the assignment based method. We hypothesize that the hard clustering done in the assignment method introduces reconstruction artifacts. This is corroborated by the model performance being most different for the SAR metric where center based consistently beats assignment based.

In the second experiment, we simulate two point sources and also add an ambient sound to all of the channels. This tests whether the model can identify sources that are spatially ambiguous. The results are shown in figure \ref{fig:2s_1a_nonoise}. We include the SDR, SIR and SAR of the ambient source in the displayed violin plots. Note that the distribution and extreme values of this experiment are less spread out than in experiment one. We hypothesize this is because it is less likely that a scene with two point sources would include spatially close  sources than a scene with three. Also note that center based separation again outperforms assignment based separation.

In the third experiment, we simulate three point sources where two of the point sources are identical recordings. This experiment tests whether the model can correctly distribute components across several locations when necessary. The results of this experiment are shown in figure \ref{fig:3s_1d_nonoise}. 
Notably, the SIR of this experiment is quite good, which means the duplicated source does not interfere with the single source. 
However, the SAR is lower, indicating additional artifacts. Overall, the model successfully finds sources even if they have multiple locations. 

We found that center based separation consistently outperformed assignment based separation. We hypothesize this is due to the artifacts introduced by hard clustering in assignment based separation. However, do note that assignment based separation is significantly faster as it only requires one forward pass of the network per source while center based separation requires an iterative procedure at each STFT frame. Results are summarized in table \ref{table:mean_results}.

\begin{table}[ht]
\renewcommand{\arraystretch}{1.325}
\centering
\caption{Mean BSS Experiment Results}
\scalebox{0.88}{
\begin{tabular}{c c c c c} 
\hline
Experiment & Separation Method & SDR (dB) & SIR (dB) & SAR (dB)\\
\hline
\multirow{2}{1em}{One} & Assignment Based & 12.40 & 18.00 & 15.12 \\ \cdashline{2-5}
& Center Based & 19.74 & 23.67 & 24.20 \\
\hline
\multirow{2}{1em}{Two}& Assignment Based & 9.09 & 14.35 & 12.20 \\ \cdashline{2-5}
& Center Based & 14.64 & 19.07 & 18.41 \\
\hline
\multirow{2}{1em}{Three}& Assignment Based & 11.89 & 20.97 & 13.61 \\ \cdashline{2-5}
& Center Based & 17.72 & 25.92 & 20.92 \\
\hline
\end{tabular}}
\label{table:mean_results}
\vspace{-8mm}
\end{table}


\begin{figure}[p]
        \centering
        \includegraphics[scale=0.35]{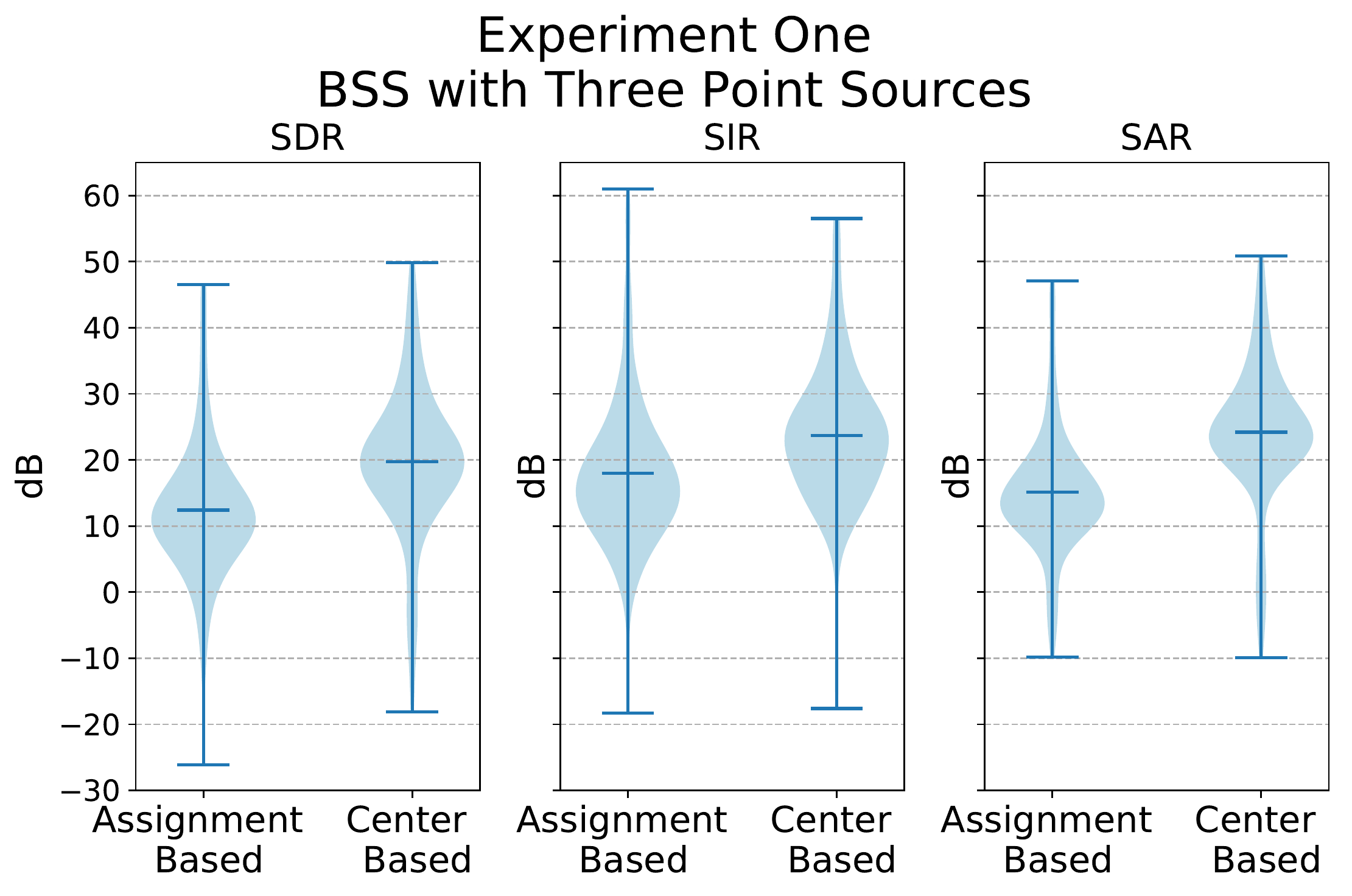}
        \vspace{-3mm}
        \caption{Results for three point sources in a reverberant room using unsupervised source separation via clustering on the channels dictionary. These plots demonstrate the model's ability to decompose scenes where each source has a distinct spatial location.}
        \label{fig:3s_nonoise}
    \vspace{.2cm}
        \centering
        \includegraphics[scale=0.35]{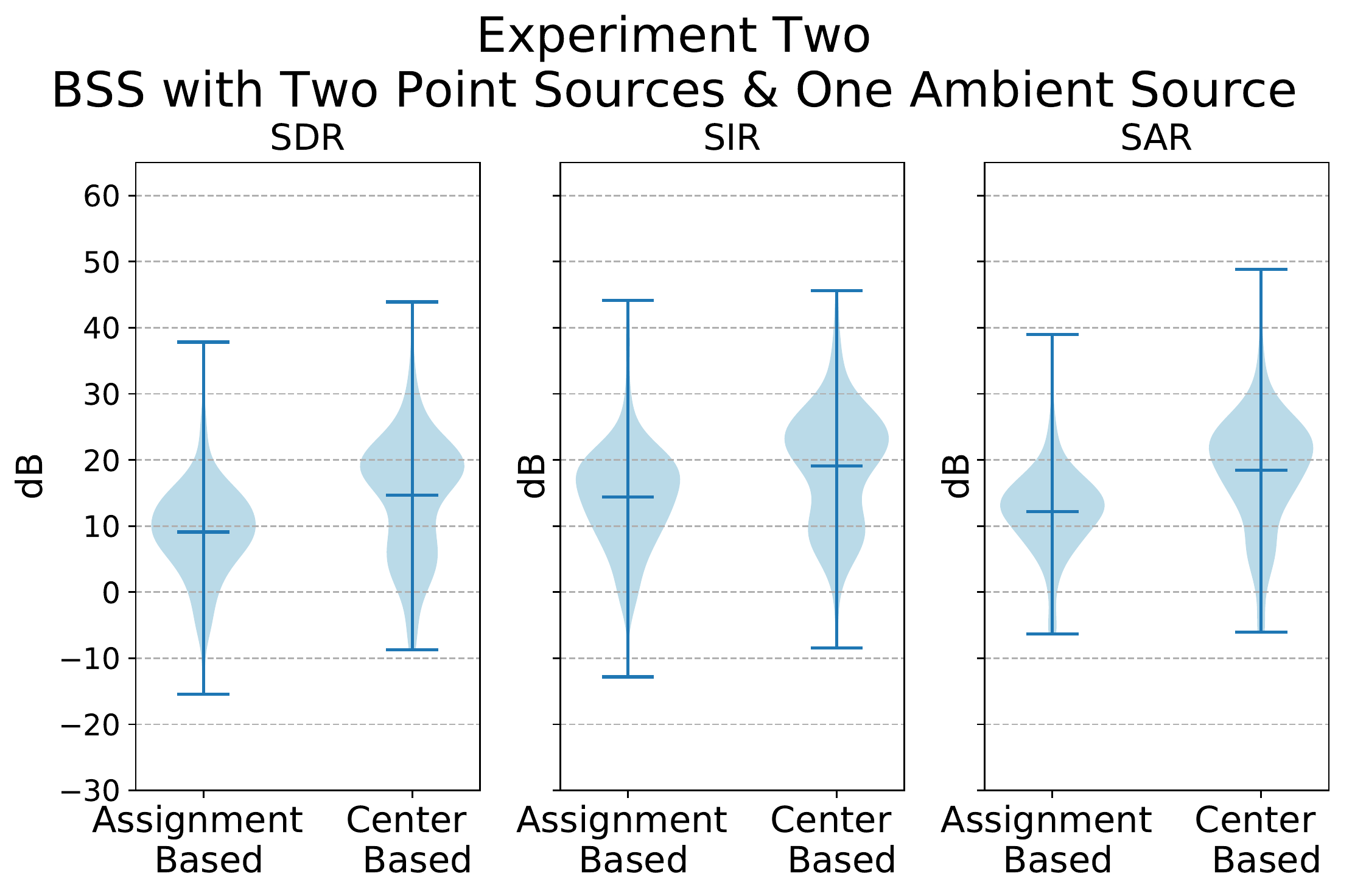}
        \vspace{-3mm}
        \caption{Results for two point sources and one ambient source in a reverberant room using unsupervised source separation via clustering on the channels dictionary. The variance in this experiment is significantly lower than experiment one. These plots demonstrate the model's ability to decompose scenes where some sources do not have a distinct spatial location.}
        \label{fig:2s_1a_nonoise}
    \vspace{.2cm}
        \centering
        \includegraphics[scale=0.35]{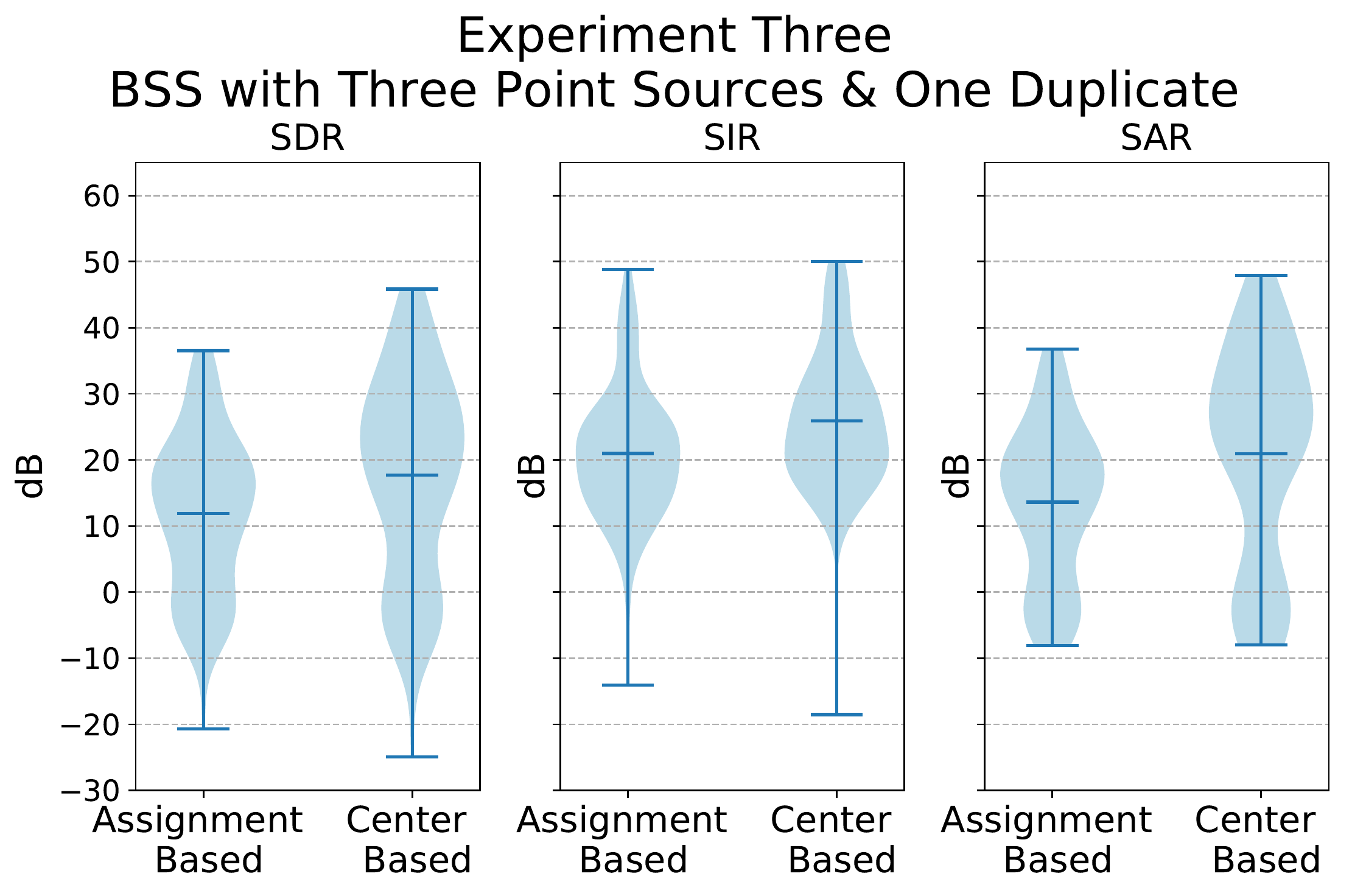}
        \vspace{-3mm}
        \caption{Results for three point sources where one channel is duplicated in a reverberant room using unsupervised source separation via clustering on the channels dictionary. The variance is higher than both previous experiments since the duplicated source negatively impacted the separations SAR. These plots demonstrate the model's ability to decompose scenes where some sources have multiple locations.}
        \label{fig:3s_1d_nonoise}
\end{figure}


\section{Conclusion}
\label{sec:conclusion}
We have presented a deep tensor factorization capable of decomposing reverberant audio scenes made up of many microphones and multiple sources. The learned decomposition can be leveraged for many downstream tasks like scene remixing and both single and multi-channel source separation. We demonstrate the utility of our decomposition on the task of blind source separation. Using our unsupervised decomposition, we perform unsupervised clustering and identify the audio sources that made up our scene. The experiments examine both point sources and ambient sources showing that our decomposition is a general-purpose model suited to many kinds of acoustics scenes. Additionally, this model can be extended to use many of the recent deep learning advances, since it is formulated as an autoencoder. For example, one can use a more complex encoder and decoder or even formulate an end-to-end model. This method can also be retro-fitted to emulate different kinds of tensor factorizations as deemed necessary from the mixing model. We hope that this model will serve as a basis for future work in unsupervised audio scene understanding by leveraging both deep methods and tensor factorizations.

\bibliographystyle{IEEEtran}
\bibliography{refs19}

\end{sloppy}
\end{document}